\documentclass[aps,pra,twocolumn,showpacs]{revtex4-1}

\usepackage{indentfirst}
\usepackage{bm}
\usepackage{verbatim}
\usepackage{graphicx}
\usepackage{subfigure}
\usepackage{epsfig}
\usepackage{amsmath}
\usepackage{booktabs}
\usepackage{multirow}
\usepackage{txfonts}
\usepackage{float}
\usepackage{tabularx}
\usepackage{threeparttable}

\usepackage[citecolor=blue,colorlinks=true,linkcolor=blue,urlcolor=blue,breaklinks=true]{hyperref}

\begin{document}

\title{Quantum Signatures of Chaos in Anisotropic Quantum Rabi Model}

\author{Shangyun Wang$^{1}$}
\author{Songbai Chen$^{2,3}$}
\author{Jiliang Jing$^{2,3}$}
\author{Jieci Wang$^{2}$}
\author{Heng Fan$^{4}$}

\email[E-mail address: ]{csb3752@hunnu.edu.cn}
\email[E-mail address: ]{jcwang@hunnu.edu.cn}
\affiliation{$^{1}$College of Physics and Electronic Engineering, Hengyang Normal University, Hengyang 421002, China \\
$^{2}$Key Laboratory of Low-Dimensional Quantum Structures and Quantum Control of Ministry of Education, Key Laboratory for Matter Microstructure and Function of Hunan Province, Department of Physics and Synergetic Innovation Center for Quantum Effects and Applications, Hunan Normal University, Changsha 410081, China \\
$^{3}$Center for Gravitation and Cosmology, College of Physical Science and Technology, Yangzhou University, Yangzhou 225009, People's Republic of China \\
$^{4}$Beijing National Laboratory for Condensed Matter Physics, Institute of Physics, Chinese Academy of Sciences, Beijing 100190, China}

\begin{abstract}
Quantum chaos is an intriguing topic and has attracted a great deal of interests in quantum mechanics and black hole physics.
Recently, the exponential growth of out-of-time-ordered correlator (OTOC) has been proposed to diagnose quantum chaos and verify the correspondence principle.
Here, we find good correspondence between the linear entanglement entropy and the semiclassical phase space structures in the anisotropic quantum Rabi model. The Loschmidt echo in the chaotic sea decays more faster than that in the stable island. However, the OTOCs grow exponentially at early times for the initial states centered both in the chaotic and stable regions. We attribute the exponential growth of the OTOC to quantum collapse which provides a novel mechanism of yielding exponential growth of the OTOC in quantum systems.
Moreover, the quantum collapse effect is more obvious for the initial states centered in the chaotic one. Our results show that in the anisotropic quantum Rabi model, the linear entanglement entropy and Loschmidt echo are more effective than OTOC for diagnosing quantum chaotic signals.

\end{abstract}
\maketitle

\section{Introduction}
The essential feature of classical chaos is its hypersensitivity to initial
conditions, i.e., the so-called butterfly effect. The tiny difference in two nearby orbits grows exponentially with time so that their motions become drastically different. Due to the uncertainty principle, there is no universal quantum counterpart of classical phase-space trajectories, so numerous methods of detecting classical chaos are invalid under quantum circumstances.
Because chaos is inherently a dynamical phenomenon, dynamical signatures such as Loschmidt echo~\cite{leo1,leo2,leo3,leo4,leo5}, entanglement entropy~\cite{EE1,EE2,EE3,EE4,EE5,EE6} and Husimi quasi-probability function~\cite{hd1,hd2,hd3,hd4} show clear differences between chaotic and regular regimes, for which have been verified both theoretically and experimentally.

In recent years, the OTOC~\cite{definition} is widely believed to be a powerful tool to detect signatures of chaos in  quantum systems because it presents an exponential growth in time $C(t) \sim e^{\lambda t}$ (where $\lambda$ is the exponential growth rate of the OTOC before the Ehrenfest time $t_{*}=\frac{1}{\lambda}\ln N$) as the corresponding classical system is in the chaotic state. Therefore, the OTOC has been used to diagnose the chaotic characteristics in various quantum systems~\cite{ot1,ot2,ot3,ot4,ot5,ot6,ot7,ot8,ot9,ot10,ot11,ot12,ot13,ot14,ot15,ot16,ot17}.

However, besides in the chaotic case, the OTOCs in some non-chaotic regular systems were also found to possess the exponential
growth behavior in early time, which means that the exponential growth of the OTOC does not imply the occurrence of chaos in quantum systems.
Thus, it is vitally important to study the mechanism of yielding the exponential growth of the OTOC.
In the integrable systems, it is shown that quantum mechanics can bring chaos to classical nonchaotic systems~\cite{sp1} and the OTOC at the saddle points grow exponentially~\cite{sp2,sp3,sp4}. Moreover, the conclusion of Ref. ~\cite{sp5} indicates that OTOCs for initial states located in unstable manifold also exhibit exponential growth behavior before the Ehrenfest time.

In this paper, we focus on the anisotropic quantum Rabi model~\cite{aqrm1,aqrm2}, which describes a two-level system coupled to a cavity electromagnetic mode. Recently, due to its well-controlled rotational and antirotational interactions, the anisotropic quantum Rabi model has attracted wide attention ranging from quantum phase transitions~\cite{aqrmqpt1,aqrmqpt2,aqrmqpt3,aqrmqpt4} to quantum state engineering~\cite{aqrm1,qse1}.
However, quantum chaos has been rarely studied in the anisotropic quantum Rabi model because it is generally regarded to be far from the so-called thermodynamic limit. If the ratio of the cavity field frequency $\omega_{0}$ to the atomic transition frequency $\omega$
approaches zero, i.e., $\omega_{0}/\omega\rightarrow 0$, the situation is changed because the truncated photon number $N_{p}$ of cavity field in this limit tends to infinity, so the system can be regarded as an effective many-body system and the thermodynamic limit~\cite{tl} can be achieved. On the other hand, the quantum collapse and revival effects have been interpreted theoretically~\cite{husimicrthe1,husimicrthe2,husimicrthe3,husimicrthe4,husimicrthe5} and observed experimentally~\cite{husimicrexp1,husimicrexp2,husimicrexp3,husimicrexp4}.
The quantum collapse and revival effects describe the splitting and merging quantum dynamical behavior of wave packets, which have no classical dynamical counterpart. In the anisotropic quantum Rabi model, Rabi oscillations lead the quantum collapse and revival effects emergence~\cite{RSAQRM1,RSAQRM2}. Then, it is natural to ask whether quantum collapse and revival effects break the classical quantum correspondence and make the quantum signature of chaos ineffective in this model? Especially, the splitting and merging behavior of wave packets in phase space is directly related to the fluctuations of coordinate and momentum operators, which is equivalent to OTOC in some physical conditions.
Here, we study the chaotic signatures in the anisotropic quantum Rabi model and find that quantum collapse provides a new mechanism for the exponential growth of the OTOC.

The paper is organized as follows. In section~\ref{SecMod}, we briefly introduce
the anisotropic quantum Rabi model and obtain its semiclassical phase space. In section~\ref{TirMod}, we study the quantum signatures of chaos in the anisotropic quantum Rabi model. In section~\ref{FouMod}, we discuss the early time evolution behavior of OTOC and propose another mechanism for the exponential growth of OTOC. Finally, we present results and a brief summary.

\section{Model and semiclassical phase space}\label{SecMod}
The Hamiltonian of the anisotropic quantum Rabi model reads ($\hbar = 1$ hereafter)~\cite{aqrm1,aqrm2}
\begin{eqnarray}
\hat{H}&=& \frac{\omega}{2} \hat{\sigma}_{z} + \omega_0 \hat{a}^{\dagger}\hat{a} + g_1(\hat{a}^{\dagger}\hat{\sigma}_{-} + \hat{a} \hat{\sigma}_{+}) \notag \\
 &+& g_2(\hat{a}^{\dagger}\hat{\sigma}_{+} + \hat{a} \hat{\sigma}_{-}),\label{ha1}
\end{eqnarray}
where $\omega$ is the atomic transition frequency, $\hat{a}^{\dag}(\hat{a})$ is the photonic creation (annihilation) operator of single mode cavity field with
frequency $\omega_0$, $g_1$ and $g_2$ are the rotating-wave and counter-rotating-wave coupling constants, respectively. $\hat{\sigma}_{\pm}=(\hat{\sigma}_x\pm i\hat{\sigma}_y)/2$ and $\hat{\sigma}_{x},\hat{\sigma}_{y},\hat{\sigma}_{z}$ are the Pauli matrices.

Here, we take the initial quantum states as
\begin{eqnarray}
|\psi(0)\rangle &=& |\tau\rangle\otimes|\beta\rangle \equiv |\tau\beta\rangle,\label{init1}
\end{eqnarray}
with
\begin{eqnarray}
|\tau\rangle &=& (1+\tau\tau^{*})^{-\frac{1}{2}}e^{\tau \hat{\sigma}_{+}}|\frac{1}{2},-\frac{1}{2}\rangle,\label{acoh1}\\
|\beta\rangle &=& e^{-\beta\beta^{*}/2}e^{\beta \hat{a}^{\dagger}}|0\rangle, \label{bcoh1}
\end{eqnarray}
and
\begin{eqnarray}
\tau &=& \frac{q_1 + ip_1}{\sqrt{2-q^{2}_1 -p^{2}_1}},  \ \ \ \ \ \beta = (q_2 + ip_2)/\sqrt{2},   \label{tb1}
\end{eqnarray}
where $|\tau\rangle$ and $|\beta\rangle$ are Bloch coherent states of atom and Glauber coherent states of bosons \cite{EE1,rmp1}, and the states $|\frac{1}{2},-\frac{1}{2}\rangle$ and $|0\rangle$ are the ground state of two-level atom and the vacuum state of cavity field, respectively.
The quantities $q_2$ and $p_2$ are the usual Cartesian coordinates of a harmonic oscillator.
For Bloch coherent state,
\begin{eqnarray}
\hat{\sigma}_{x} &=& j\sin\theta\cos\varphi,\\
\hat{\sigma}_{y} &=& j\sin\theta\sin\varphi,\\
\hat{\sigma}_{z} &=& j\cos\theta,
\end{eqnarray}
where $j$, $\theta$ and $\varphi$ are the spin length, polar and azimuthal angle, respectively.
Note that $\hat{\sigma}_{x}$ and $\hat{\sigma}_{y}$ do not form a pair of canonical variables.
Making a change of variables to go from action-angle to Cartesian coordinates\cite{mf}, we have
\begin{eqnarray}
q_1 &=& \sqrt{2(j+j\cos \theta)}\sin \varphi
\end{eqnarray}
and
\begin{eqnarray}
p_1 &=& \sqrt{2(j+j\cos \theta)} \cos \varphi.
\end{eqnarray}
The variables $q_1$ and $p_1$ do not exhibit direct physical interpretation and are associated with the projections $\hat{\sigma}_{x}$ and $\hat{\sigma}_{y} $ by $\hat{\sigma}_{y}/\hat{\sigma}_{x} = q_1/p_1$.
The selection of initial quantum states are based on their wave packets are minimal in phase space. With the mean field approximation procedure,
\begin{eqnarray}
\langle\tau|\hat{\sigma}_{+}|\tau\rangle &=& \frac{\tau^*}{1+\tau\tau^*},
\end{eqnarray}
\begin{eqnarray}
\langle\tau|\hat{\sigma}_{-}|\tau\rangle &=& \frac{\tau}{1+\tau\tau^*},
\end{eqnarray}
\begin{eqnarray}
\langle\tau|\hat{\sigma}_{z}|\tau\rangle &=& -\frac{1-\tau\tau^*}{1+\tau\tau^*},
\end{eqnarray}
\begin{eqnarray}
\langle\beta|\hat{a}|\beta\rangle &=& \beta,
\end{eqnarray}
\begin{eqnarray}
\langle\beta|\hat{a}^{\dagger}|\beta\rangle &=& \beta^*,
\end{eqnarray}
one can obtain the semiclassical Hamiltonian related to the anisotropic quantum Rabi model,
\begin{eqnarray}
H_{cl} &\equiv& \langle \tau\beta|\hat{H}|\tau\beta\rangle = \frac{\omega}{2} (q^{2}_1 +p^{2}_1 -1)+ \frac{\omega_0}{2}(q^{2}_2 + p^{2}_2) \notag \\
&+&\sqrt{1-(q^{2}_1 +p^{2}_1)/2}( G_+ q_1 q_2 + G_- p_1 p_2),\label{ha2}
\end{eqnarray}
where $G_{\pm} = g_1 \pm g_2$.
\begin{figure}[ht]
\begin{center}
\includegraphics[width=6cm]{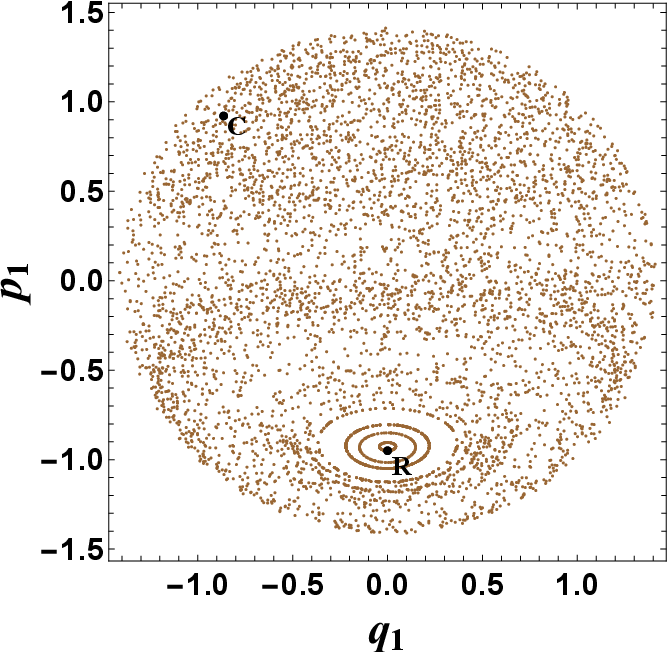}
\caption{The Poincar\'{e} section for the anisotropic quantum Rabi model in the case: $q_{2}=0, p_{2}>0$, with $\omega=1$, $\omega_{0} =0.2$, $g_1=0.9$, $g_2=0.5$ and the system energy $E = 2$. Points $R (q_{1} = 0,p_{1} =-0.95,q_{2} =0,p_{2} =6.14757)$ and $C (q_{1} = -0.86413,p_{1} =0.92136,q_{2} =0,p_{2} =3.37955)$ are situated in stable island and chaotic sea respectively.}\label{f1}
\end{center}
\end{figure}
With the classical Hamiltonian~\ref{ha2}, we obtain a poicar\'{e} section of the anisotropic quantum Rabi model, as shown in Fig. \ref{f1}.
For this section, resulting in a mixed phase space that contains both regular and chaotic regions composed of many discrete points. Motion across the boundaries between regular and chaotic regions is classically forbidden. Here, the classical Lyapunov exponent $\Lambda$ at the initial point $C$ in the chaotic regions is $\Lambda_{C} \approx 0.134$. For the point $R$ in the regular region, its classical Lyapunov exponent is $\Lambda_{R} \approx 0$ (see Fig. S1, Supporting Information). Especially, when $g_2$ equal to 0, the Hamiltonian~\ref{ha2} becomes integrable and no chaotic orbits appear in the anisotropic quantum Rabi model.

\section{Quantum signatures of chaos}\label{TirMod}
\subsection{Linear Entanglement Entropy}
\begin{figure}[ht]
\begin{center}
\includegraphics[width=6cm]{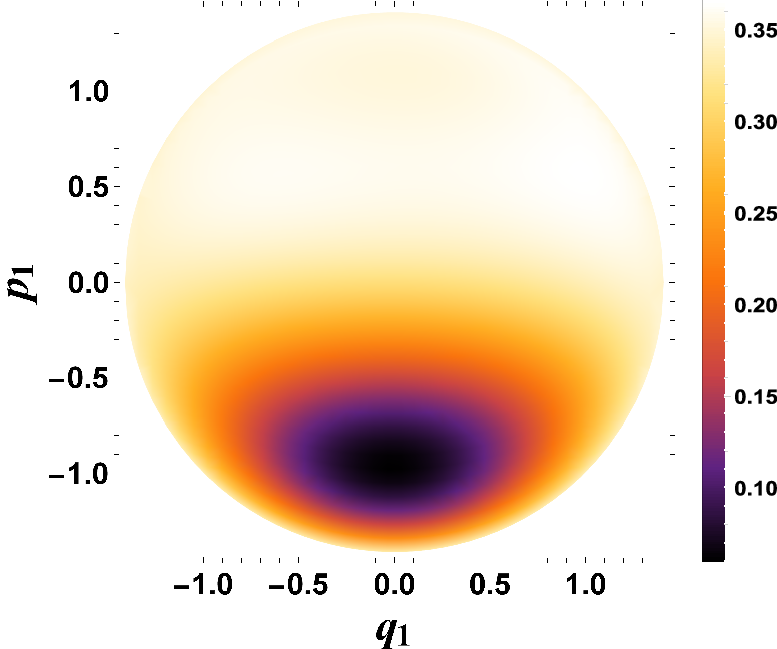}
\caption{The distribution of time-average entanglement entropy $S_{m}$ of Poincar\'{e} section in Fig. \ref{f1},
where the integral interval is $t\in[0, 50]$ and the photon number is truncated at $N_p=150 $ (see Fig. S2, Supporting Information).}\label{f2}
\end{center}
\end{figure}

The entanglement entropy has been widely used to diagnose chaos in quantum systems. In the semiclassical regime, since the dynamics of the entanglement entropy is governed by the divergence of nearby phase-space trajectories via the local Lyapunov spectrum, and thus more entanglement generation for initial states localized in chaotic regions compared to those in regular regions~\cite{EE3,EE6}. The linear entanglement entropy is one of easily measurable entropies in experiments~\cite{EE3}. It is defined as
\begin{eqnarray}
S(t) &=& 1- \rm Tr_{2} \hat{\rho}_{2}(t)^2,   \label{cha1}
\end{eqnarray}
with the reduced density matrix
\begin{eqnarray}
\hat{\rho}_{2} &=& \rm Tr_{1}|\psi(t)\rangle\langle\psi(t)|,   \label{cha2}
\end{eqnarray}
where $\rm Tr_{i}$ is a trace over the $i$ th subsystem($i=1,2$) and the wave function $|\psi(t)\rangle$ is the quantum state of the system.
Figure \ref{f2} shows that the distribution of the time-average entanglement entropy  $S_{m}=\frac{1}{T}\int_{t_{1}}^{t_{2}}S(t)dt$ in phase space has a good correspondence with the classical poicar\'{e} section in Fig. \ref{f1}. The significant dip in the time-average entanglement entropy between chaotic and regular regions means that the entanglement entropy can well diagnose the quantum chaos in the anisotropic quantum Rabi model.
It indicates that the classical quantum correspondence can be realized in the system for  single atom interact with cavity field.
The correspondence between the time-average entanglement entropy and the semiclassical phase space provides a direct evidence for defining the semiclassical limit of quantum Rabi Hamiltonian~\cite{dsrabi}. Furthermore, it also provides a specific and feasible way for studying the semiclassical dynamics in the anisotropic quantum Rabi model.
\subsection{Loschmidt echo}
The Loschmidt echo is another effective way to diagnose  the signals of chaos in quantum systems.
Considering two Hamiltonians with slight differences acting on the initial state $\psi(0)$, the Loschmidt echo can describe the sensitivity in quantum dynamics similar to the classical dynamics. Let us assume that there exists a constant perturbation $\delta$ in the atomic transition frequency in the Hamiltonian,
and then the Loschmidt echo can be expressed as
\begin{eqnarray}
L(t) &=& |\langle\psi(0)|e^{i\hat{H}(\omega)t}e^{-i\hat{H}'(\omega + \delta)t}|\psi(0)\rangle|^2,   \label{lt1}
\end{eqnarray}
where $H'(\omega + \delta)$ is the Hamiltonian after a perturbation parameter $\delta$ acts on the atomic frequency $\omega$.
According to the quantum theory, the overlap of two initial wave functions is supposed
to be $1$. With the evolution of the two states under the action of two Hamiltonians,
the Loschmidt echo decays exponentially for the initial state centered in the chaotic region, but decays slowly for that in the stable one~\cite{leo1,leo2,leo3,leo4,leo5}.
\begin{figure}[ht]
\begin{center}
\includegraphics[width=6cm]{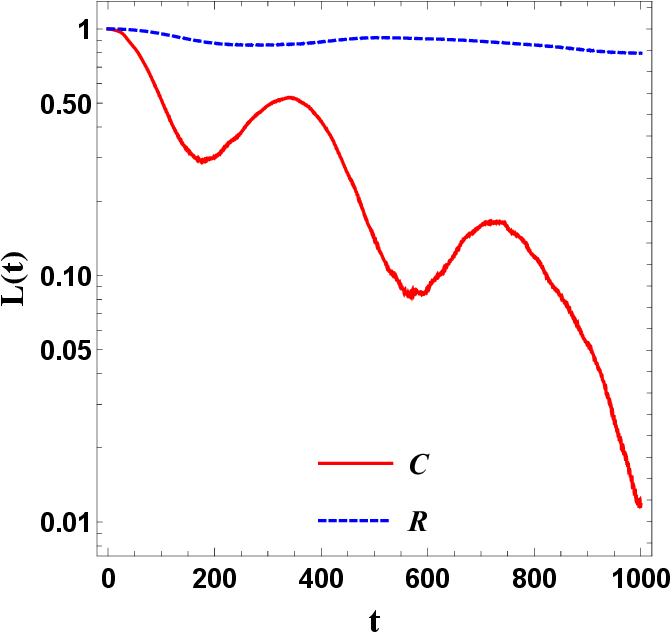}
\caption{
The Loschmidt echo $L(t)$ is computed for coherent states centered at the chaotic point $C$ and regular point $R$ in Fig. \ref{f1}.
Here, we set $\delta=0.1$, $\omega=1$, $\omega_{0}=0.2$, $g_1=0.9$, $g_2=0.5$ and the photon number is truncated at $N_p=150$.}\label{f3}
\end{center}
\end{figure}
In Fig. \ref{f3}, we present the evolution of Loschmidt echo for the initial states centered at the point $C$ (in the chaotic region) and $R$ (in the stable region). Obviously, the Loschmidt echo decays oscillatorily with a larger exponent in the case of the point $C$, and decays more slowly in the case of the point $R$. Therefore, the Loschmidt echo clearly identifies the signature of quantum chaos in the anisotropic quantum Rabi model.
\begin{figure}[ht]
\begin{center}
\includegraphics[width=6.5cm]{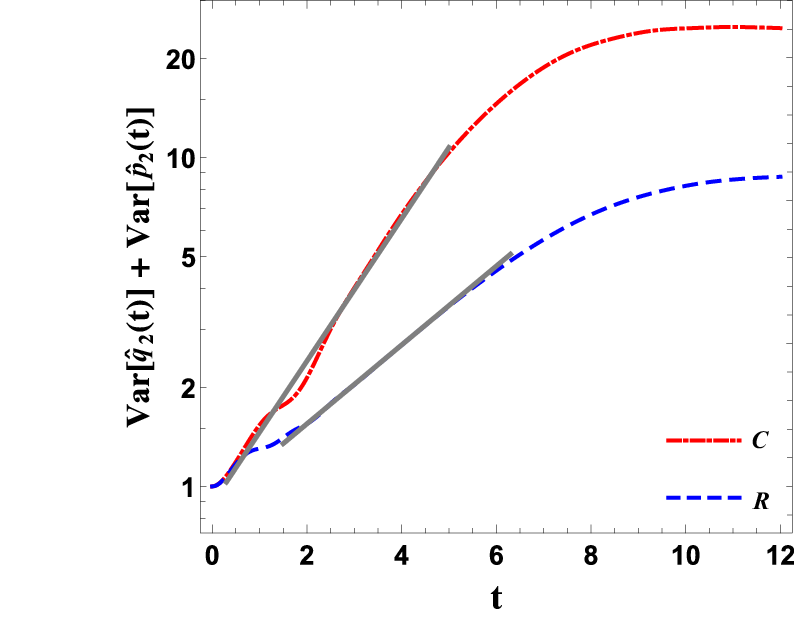}
\caption{Time evolution of the OTOC $Var[\hat{q}_2(t)] + Var[\hat{p}_2(t)]$ for the initial coherent states centered at chaotic point $C$ and regular point $R$.
The gray line is the fitted line and the exponential growth rate of points C and R are $\lambda_C\approx0.498$ and $\lambda_R\approx0.276$, respectively.
Here, we set $\omega=1$, $\omega_{0}=0.2$, $g_1=0.9$, $g_2=0.5$ and the photon number is truncated at $N_p=150$.}\label{f4}
\end{center}
\end{figure}
\begin{figure*}[htb]
\centering
\epsfysize=6cm \epsfclipoff \fboxsep=0pt
\setlength{\unitlength}{1.cm}
\begin{picture}(12,5)(0,0)
\put(-2,0.0){{\epsffile{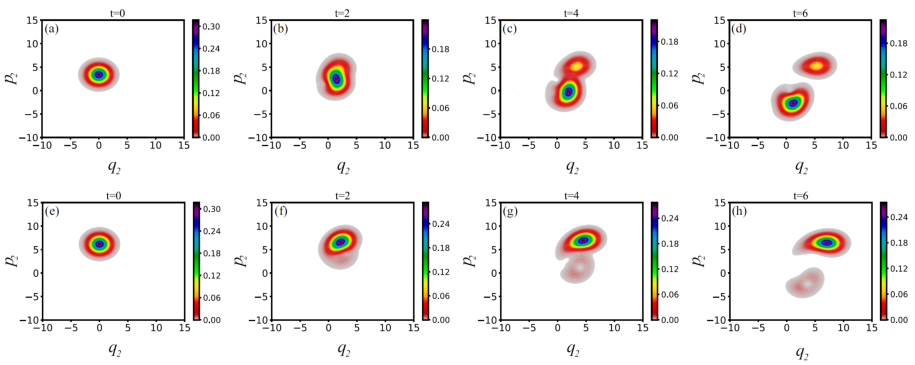}}}
\end{picture}
\caption{ The time evolution of the Husimi quasi-probabilistic wave packets.
The top and bottom panels denote respectively the case in which the initial coherent states centered at the points
$C$ and $R$ in Fig. \ref{f1}. Here, we set $\omega=1$, $\omega_{0}=0.2$, $g_1=0.9$, $g_2=0.5$ and the photon number is truncated at $N_p=150$. }
\label{f5}
\end{figure*}
\section{out-of-Time-Ordered Correlator and Quantum Collapse}\label{FouMod}
Let us now focus on the OTOC, which is another indicator to detect quantum chaos
and its form can be defined as~\cite{ot1,definition}
\begin{eqnarray}
C(t) &=& -\langle[\hat{V}(0),\hat{W}(t)]^2\rangle,   \label{otoc1}
\end{eqnarray}
where $\langle...\rangle$ denotes the expectation values and $\hat{W}(t)=e^{i\hat{H}t}\hat{W}e^{-i\hat{H}t}$. $\hat{H}$ is a quantum Hamiltonian,
$\hat{W}$ and $\hat{V}$ are two approximately local operators~\cite{ot1}.
Here, we set $\hat{V}=|\psi(0)\rangle\langle\psi(0)|$ as the projection operator onto the initial state, and the OTOC (\ref{otoc1}) becomes
\begin{align}
C(t)=& \langle\psi(0)| \hat{V}\hat{W}(t)\hat{W}(t)\hat{V}|\psi(0)\rangle+\langle\psi(0)| \hat{W}(t)\hat{V}\hat{V}\hat{W}(t)|\psi(0)\rangle\nonumber\\
-&\langle\psi(0)| \hat{V}\hat{W}(t)\hat{V}\hat{W}(t)|\psi(0)\rangle-\langle\psi(0)| \hat{W}(t)\hat{V}\hat{W}(t)\hat{V}|\psi(0)\rangle\nonumber\\
=&\langle\psi(t)|\hat{W}^{2}|\psi(t)\rangle - \langle\psi(t)|\hat{W}|\psi(t)\rangle^{2}\nonumber\\
\equiv& Var[\hat{W}(t)],  \label{otoc2}
\end{align}
where $Var[\hat{W}(t)]$ is the quantum variance of operator $\hat{W}$. This relation establishes a connection between quantum variances and quantum chaos, and enables us to
visualize the evolution in terms of the quantum dynamics using well-known phase-space methods, such as Husimi Q function. Moreover, it measures the spread of the size of the wave packet when the operator $\hat{W}$ was selected as coordinate and momentum operators.

Since the wave packet of the cavity field spreads in both directions in phase space, we plot the OTOC (\ref{otoc2}) as $Var[\hat{q}_2(t)] + Var[\hat{p}_2(t)]$ for the anisotropic quantum Rabi model with different initial states. Here, $\hat{q}_2=(\hat{a}^{\dag} + \hat{a})/\sqrt{2}$ and $\hat{p}_2= i(\hat{a}^{\dag} - \hat{a})/\sqrt{2}.$ As shown in Fig. \ref{f4},
we find that the OTOC for the initial coherent state centered at the point $R$ exhibits an exponential growth behavior, while its corresponding classical orbit is stable. For the initial coherent state centered at the point $C$ in the chaotic region, we find the OTOC still grows exponentially, and its exponential growth rate is greater than the twice of the corresponding classical Lyapunov exponent, i.e., $\lambda_C>2\Lambda_C$. This against the general relationship between the exponential growth rates of OTOCs and classical Lyapunov exponents $\lambda_C=2\Lambda_C$~\cite{ot2,ot3,ot10}. These results indicate that the exponential growth of the OTOC in the anisotropic quantum Rabi model could be caused by other unknown factors besides chaos.

To further unveil the factors yielding the exponential growth of the OTOC in the anisotropic quantum Rabi model, we analyze the evolution of Husimi Q function~\cite{husimi1} during the OTOC grows exponentially. For coherent states, the Husimi Q function is defined
as $Q(q_2,p_2)=\frac{1}{\pi} \langle q_{2},p_2|\hat{\rho}_2|q_2,p_2\rangle$, where $\rho_2$ is the reduced density matrix of the second subsystem and $|q_2,p_2\rangle$ is a photon coherent state. Husimi Q function demonstrates the dynamical evolution of the quantum state with time and provide a visualization of high-dimensional quantum states. In Fig. \ref{f5}, we show the evolution of Husimi Q function during the exponential growth of the OTOC for different initial states in the phase space of anisotropic quantum Rabi model. We find that the quantum collapse (the splitting of the wave packet)~\cite{husimicrthe5} emerges during the exponential behavior of the OTOC, which does not depend on whether the initial states centered in the chaotic sea or stable island. Compared to the dynamic evolution without quantum collapse and revival effect, the splitting of wave packet in phase space causes an additional growth in coordinate and momentum variances. This is the essence why the OTOCs in the anisotropic quantum Rabi model exhibit exponential growth for regular initial states and grow exponentially at a rate greater than twice the classical Lyapunov exponent for chaotic initial states.
\begin{figure}[ht]
\begin{center}
\includegraphics[width=6cm]{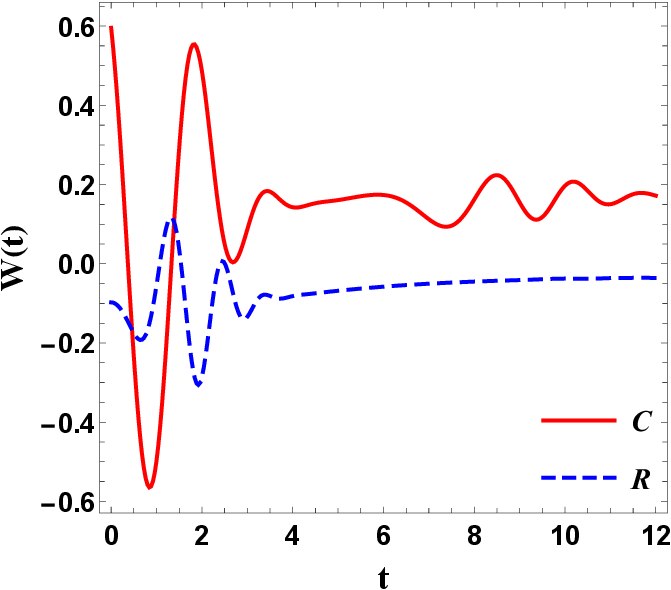}
\caption{The evolution of atomic population inversion W(t) for the initial coherent states centered at chaotic point C and regular point R. Here, we set $\omega = 1$, $\omega_{0} =0.2$, $g_1=0.9$, $g_2=0.5$ and the system energy $E=2$.}
\label{f33}
\end{center}
\end{figure}
The appearance of quantum collapses in both cases is further confirmed by the atomic population inversion with time $W(t)$ as shown in Fig. \ref{f33}.  The atomic population inversion is $W(t)\equiv P_{e}(t) - P_{g}(t)$, where $P_{e}(t)$ and $P_{g}(t)$ are atomic populations in the excited state $|1/2,1/2\rangle$ and the ground state $|1/2,-1/2\rangle$, respectively. The oscillatory phase of $W(t)$ at early time corresponding to the splitting process of wave packets in quasi-probability distribution in Fig. \ref{f5}, which means the emergence of quantum collapse of Rabi oscillation~\cite{husimicrthe5}.
The split behavior of quantum wave packet in short time implies that the quantum variances of momentum and coordinate operators increase rapidly.
This indicates that quantum collapse also gives rise to the exponential growth behavior of the OTOC in the anisotropic quantum Rabi model. To further demonstrate that quantum collapse effects can lead OTOCs grow exponentially, we show that the OTOCs in the Jaynes-Cummings (JC) model and the regular anisotropic quantum Rabi model also exhibit exponential growth behavior (see Fig. S3-S6, Supporting Information).

Comparing Fig. \ref{f5}(b) and Fig. \ref{f5}(f), Fig. \ref{f5}(c)and Fig. \ref{f5}(g), we find that the quantum collapse effects for the initial coherent states centered in the chaotic sea are more obvious than that in the regular regions. This means that the Husimi Q function can still be an effective tool for diagnosing chaos in the anisotropic quantum Rabi model.

In the Fig. S7-S11 of Supporting Information, we show that linear entanglement entropy and Loschmidt echo show more advantages than OTOC for diagnosing chaos in the two-photon Rabi model with spectral collapse effect. Moreover, we find that quantum wave packets continuously split into multiple wave packets during the OTOCs increase at an exponential rate. This further indicates that quantum collapse gives rise to the exponential growth behavior of the OTOCs.

\section{Conclusion }\label{SecCon}
Although the OTOC is generally regarded as a tool of diagnosing quantum chaos, the exponential growth of the OTOC can also be caused by non-chaotic behaviors of quantum systems. Here, we find that quantum collapse can give rise to the exponential growth behavior of the OTOC, which provides a novel mechanism of yielding exponential growth of the OTOC in quantum systems.
Moreover, we show that the classical quantum correspondence can be realized in the anisotropic quantum Rabi model, and both entanglement entropy and Loschmidt echo effectively diagnose quantum chaotic signals in this system.

The exponential growth behavior of OTOCs in regular systems are usually closely related to saddle points, as saddle points show instability similar to chaotic orbits, i.e., classical Lyapunov exponents are positive.
Essentially, quantum information is determined by wave functions.
Therefore, when the wave functions in regular systems exhibit quantum dynamical behaviors without classical dynamic counterparts, the OTOCs may exhibit false chaotic behaviors.
A foreseeable fact is that when the wave packets in regular systems split into three or more wave packets simultaneously, the OTOCs will exhibit more pronounced exponential growth behavior.\\

\section*{Acknowledgments}
This work was supported by the Science Foundation of Hengyang Normal University of China under Contract No.2020QD24; the National Natural Science Foundation of China under Grants No.12475051, No.12374408, and No.12035005; the science and technology innovation Program of Hunan Province under grant No.2024RC1050; the innovative research group of Hunan Province under Grant No.2024JJ1006; and Scientifc Research Fund of Hunan Provincial Education Department(Grant No.24C0347).

\end{document}